\documentclass{article}
\usepackage{spconf,amsmath,graphicx, xltabular, booktabs}

\usepackage{enumitem}
\setlist{nosep, leftmargin=14pt}

\usepackage{mwe} 

\newcommand{\methodname}{PET-Disentangler\xspace}
\title{Disentangled PET Lesion Segmentation}

\name{Tanya Gatsak$^{1}$, Kumar Abhishek$^{1}$, Hanene Ben Yedder$^{1}$, Saeid Asgari Taghanaki$^{1, 2}$, Ghassan Hamarneh$^{1}$}

\address{ $^{1}$Medical Image Analysis Lab, School of Computing Science, Simon Fraser University, Canada \\ $^{2}$Autodesk AI Research, Canada}

\begin{document}
\maketitle
\begin{abstract} 
PET imaging is an invaluable tool in clinical settings as it captures the functional activity of both healthy anatomy and cancerous lesions. Developing automatic lesion segmentation methods for PET images is crucial since manual lesion segmentation is laborious and prone to inter- and intra-observer variability. We propose \methodname, a 3D disentanglement method that uses a 3D UNet-like encoder-decoder architecture to disentangle disease and normal healthy anatomical features with losses for segmentation, reconstruction, and healthy component plausibility. A critic network is used to encourage the healthy latent features to match the distribution of healthy samples and thus encourages these features to not contain any lesion-related features. Our quantitative results show that \methodname is less prone to incorrectly declaring healthy and high tracer uptake regions as cancerous lesions, since such uptake pattern would be assigned to the disentangled healthy component.

\end{abstract}
\begin{keywords}
Positron Emission Tomography (PET), Image Segmentation, Disentangled Representations
\end{keywords}

\section{Introduction}
\label{sec:intro}
Positron emission tomography (PET) is a medical imaging modality that captures functional, metabolic activity within the body. PET is used in many areas of medicine, especially in oncology for cancer staging, diagnosis, and monitoring \cite{beichel2017multi}. Due to the nature of the modality, areas of high metabolic activity, including both healthy anatomy and cancerous lesions, correspond to high intensity values. Expert clinicians' analysis of PET images is laborious and can suffer from inter- and intra-observer variability, neccessitating automated methods. 

Over the past few decades, automatic PET lesion segmentation methods have evolved from various intensity thresholding-based approaches, active contours, and region growing algorithms~\cite{foster2014review} to the current state-of-the-art methods that rely on deep learning (DL) to optimize models to learn disease features~\cite{yousefirizi2021toward}. 
A promising DL-based approach that has yet to be explored in PET lesion segmentation is image disentanglement, which attempts to separate distinct sources of variation within images into independent representations.
Image disentanglement has been shown to be beneficial in a wide variety of medical image analysis applications \cite{liu2022learning}. However, there is no work on evaluating the utility of an image disentanglement-based approach for PET images, a challenging modality owing to the presence of noise, low signal-to-contrast ratio, and high activity regions corresponding to both healthy and disease activity. 

In this work, we leverage lesion segmentation with a disentanglement framework and introduce \methodname, a novel PET segmentation method that disentangles 3D PET images in the latent space into disease features and normal healthy anatomical features for a robust and explainable segmentation method. \methodname uses the disease features for lesion segmentation prediction, the healthy features to estimate pseudo-healthy images per input, and re-entangles both healthy and disease features for a full reconstruction. \methodname enhances the lesion segmentation task by providing explainability in the form of a pseudo-healthy image as to what the model expects the lesion-free image to look like per given input. Furthermore, \methodname shows that learning the healthy component provides a solution to a critical challenge in PET lesion segmentation \cite{ahmadvand2016tumor}, where segmentation models can incorrectly segment healthy, high-intensity areas as disease. This challenge has seen solutions that focus on localizing healthy, high intensity regions \cite{afshari2018automatic} whereas the proposed \methodname can both capture the healthy anatomy features and delineate lesions through the learned disentangled representations.  

\section{Method}
\label{sec:method}

\begin{figure}
    \centering
    \includegraphics[width=\columnwidth]{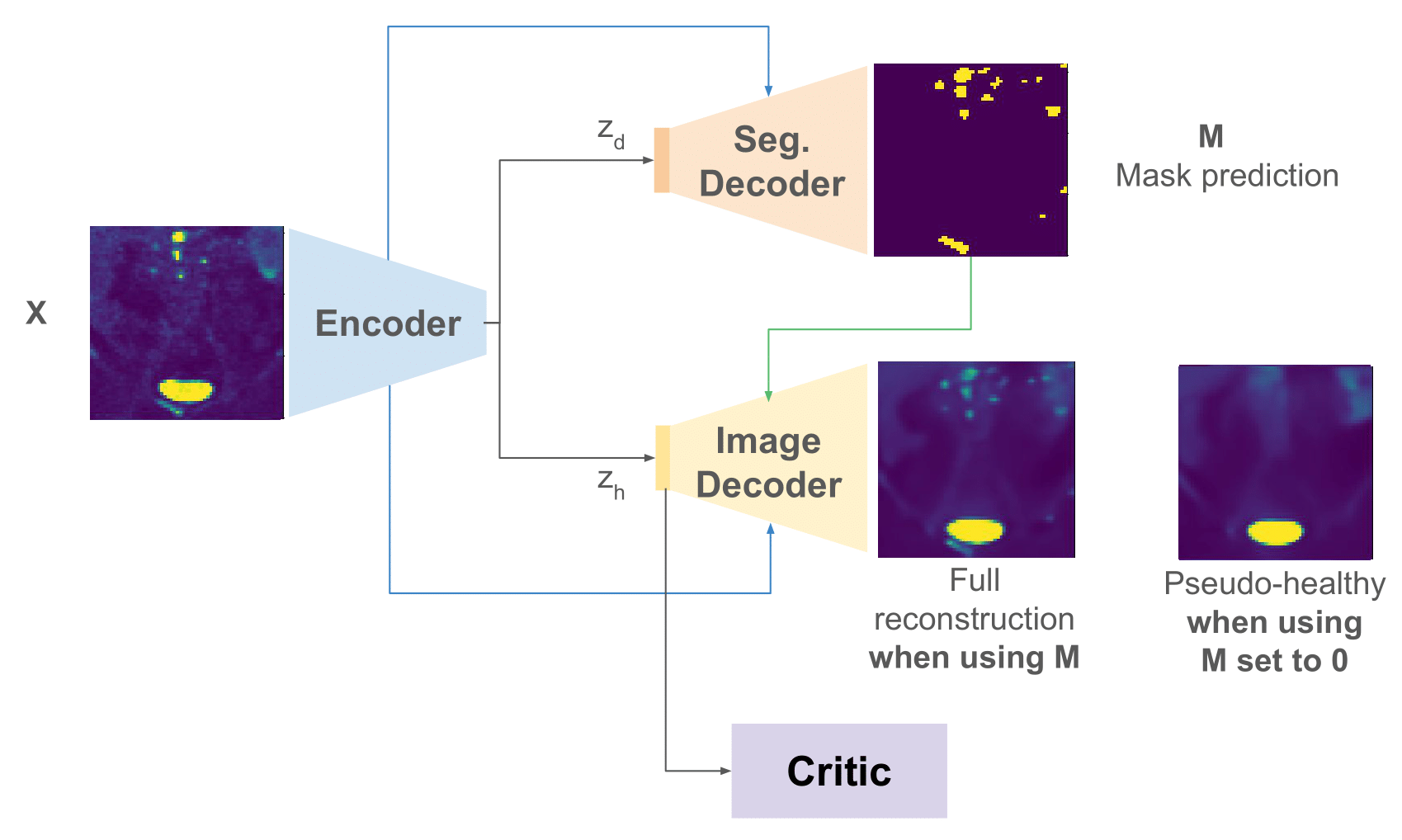}
    \caption{The proposed disentanglement architecture of \methodname facilitates the learning of disease features through two pathways: a segmentation prediction path and image reconstruction path. By re-entangling healthy and disease-specific features, this design enables \methodname to effectively capture disease characteristics while maintaining an accurate representation of healthy anatomy. The black arrows indicate feature flow throughout the network,  the blue arrows represent skip connections, and the green arrow represents the use of the mask prediction in the image decoder. }
    \label{fig:schematic-model-architecture}
\end{figure}

An overview of the proposed architecture is shown in Fig. \ref{fig:schematic-model-architecture}, including a critic network that is used to ensure that healthy features do not contain lesion features. \methodname adopts a UNet-like architecture that is extended from 2D to 3D and further modified to have one encoder and two decoders for segmentation prediction and for reconstruction. The encoder takes as input a PET image $X$ and outputs two bottleneck latent vectors $z_h$ and $z_d$, which encode the healthy and disease features of the input image, respectively. The disease features are passed to the segmentation decoder that predicts a segmentation mask $M$. The healthy features are passed along with the segmentation mask to the image decoder to re-entangle the healthy and disease features and produce a reconstruction of the input $R$. Skip connections are used between each encoder and segmentation decoder block, where the number of skip connections between the encoder and the image decoder are tuned to prevent the introduction of disease features. This is because the disease features, if any exist, should be introduced via the segmentation mask during image decoding. 

\subsection{Modified Encoder Architecture}
The encoder takes in a 3D PET image and applies a series of encoding blocks, consisting of a 3D convolution, batch normalization and ReLU, to produce the latent vectors $z_h$ and $z_d$. The first encoding block takes $X$ as input and produces an output vector that is then split into healthy and disease features via the application of two separate encoding blocks. For the encoding, separate encoding blocks are applied to healthy and disease features.

\subsection{Segmentation Decoder and Prediction}
The disease features $z_d$ are passed to the segmentation decoder with the corresponding skip connections to generate a segmentation mask prediction $M$. Each decoding block consists of a transposed 3D convolution to upsample the spatial dimensions of the feature vector, concatenation between the feature vector and skip connection from the encoder, followed by a sequence of 3D convolution, batch normalization and ReLU. 

To optimize the segmentation mask prediction, ComboLoss \cite{taghanaki2019combo}, $L_{Combo}$, is used between the predicted mask and the ground truth mask, $M_{GT}$: 
\begin{equation}
    L_{seg} = L_{Combo}(M, M_{GT}) = L_{Dice} + L_{CE},
\end{equation} 
where $M_{GT}$ is binary such that voxels labelled as 0 belong to either background or healthy anatomy whereas voxels labelled as 1 belong to cancerous lesions.  

\subsection{Image Decoder and Reconstruction}
The image decoder uses the healthy features $z_h$ and corresponding skip connections along with the segmentation prediction $M$ during decoding to generate image reconstructions. When the segmentation mask prediction contains lesion features, the image decoder is optimized to predict $R$, the full reconstruction of input $X$. When the mask prediction is set to empty such that it contains all zeros, $M_0$, the image decoder is optimized to leverage only the healthy features to generate $P$, a pseudo-healthy estimate of the input $X$. For inputs $X$ that have no tumours, $R$ and $P$ should be identical.

The image decoder re-entangles the healthy and disease features using spatially-adaptive normalization (SPADE) blocks \cite{park2019semantic} to combine the healthy features at each resolution during image decoding with a downsampled version of the segmentation mask of the same spatial dimensions. The image decoder follows a similar architecture as the segmentation decoder, with the key difference being the introduction of SPADE blocks to combine healthy and disease features. Another difference is that the skip connections are removed from the last three decoding blocks to prevent disease features from potentially being introduced from outside of the mask prediction. 

To optimize the model to learn the reconstruction $R$, $L_1$ and $L_2$ reconstruction losses are used: \\
\begin{equation}
    L_{recon} = ||X-R||_1 + ||X-R||_2.
\end{equation} 

\subsection{Critic Network for Healthy Distribution Matching}
\noindent To ensure that $z_h$ only contains features relating to healthy anatomy, regardless of whether the input image has lesions, a Wasserstein GAN (WGAN) is used as a critic network with gradient penalty to align healthy features to a healthy distribution \cite{kobayashi2021decomposing}.

A set of images can be partitioned into those without any tumour lesions (i.e., negative findings) $X^-$ and those with tumour lesions (i.e., positive findings) $X^+$, where the corresponding healthy feature vectors are $z^-_h$ and $z^+_h$, respectively. Ideally, the healthy feature vectors should only contain features for healthy anatomy regardless of whether the input image has disease, and the distributions corresponding to the sets of healthy feature vectors should match. The process of distribution matching will eliminate any leakage of disease in $z_h$, as disease features will be outside of the distribution of healthy anatomy, thus ensuring that $z_h$ from all examples only contain healthy features. The critic network is optimized using the $L_{critic}$ loss:
\begin{equation}
    \begin{split}
        L_{critic} &=  w_c\left( - \left( C(z_h^-) - C(z_h^+) \right) \right. \\
        &\quad + \left. \lambda_{GP} \left( ||\nabla_{z_m} C(z_m)||_2 - 1 \right)^2 \right),
    \end{split}
\end{equation}

\noindent where the two terms correspond to the Wasserstein distance and the gradient penalty, respectively. The gradient penalty uses an interpolated healthy vector $z_m$ between $z^-_h$ and $z^+_h$, that is weighted by   $\alpha$ and calculated as follows:
\begin{equation}
    z_m = \alpha z^-_h + (1-\alpha)z^+_h.
\end{equation}

As the critic network learns to identify the difference between $z^-_h$ and $z^+_h$, the encoder tries to produce healthy feature vectors that appear from the same distribution. As such, the corresponding WGAN term used in the overall loss is referred to as the pseudo-healthy loss $L_{pseudo{\text -}healthy}$, described by:
\begin{equation}
    L_{pseudo{\text -}healthy} = -C(z_h^+).
\end{equation}

\subsection{Overall Objective Function}
The critic network is optimized separately from the rest of the network components, whereas the overall objective function to optimize the encoder, segmentation decoder, and image decoder is: 
\begin{equation}
    L_{overall} = w_s\:L_{seg} + w_r\:L_{recon} + w_{ph}\:L_{pseudo{\text -}healthy}.
\end{equation} 
The parameters $w_s$, $w_r$, and $w_{ph}$ refer to the weights of the contribution for the segmentation, reconstruction, and pseudo-healthy losses to the overall loss, $L_{overall}$.

\section{Experiments}
\label{sec:expperiments}
\noindent \textbf{Dataset:} We evaluate \methodname on TCIA whole-body FDG-PET/CT dataset \cite{gatidis2022whole} that consists of 900 patients and a total of 1014 scans, where 513 scans have no cancerous lesions and 501 scans have lesions from either lymphoma, melanoma, or lung cancer. As a core contribution of our method is the ability to model and learn the healthy anatomy component to ultimately distinguish disease features when they are present, we require volumes of the same relative area between all healthy and disease examples to learn from. TotalSegmentator \cite{wasserthal2023totalsegmentator} is leveraged to obtain anatomical segmentations for each whole-body PET-aligned CT scan, in which these segmentations act as landmarks to center spatial croppings in the PET scans and obtain a set of aligned subvolumes for the dataset. We utilize the bladder segmentation to center a lower torso $128\times128\times128$ cropped PET volume that is further resized to $64\times64\times64$ using bilinear interpolation. We then clip PET SUV values to [0, 15] and normalize to [0, 1]. The healthy and disease examples are split into 80:10:10 splits for training, validation, and testing sets. \\
\noindent \textbf{Implementation Details:}
\methodname was implemented using PyTorch and MONAI. We used the Adam optimizer with a learning rate of 1e-3 and a batch size of 4 volumes of $64^3$ voxels, consisting of 2 healthy and 2 disease examples per batch, to train the models. Each model was trained for 300 epochs using a NVIDIA A5000 GPU (24 GB) and the model with the lowest $L_{Combo}$ on the validation set was used for evaluation. The weights $w_s$, $w_r$, $w_{ph}$, and $w_c$ were empirically found to be 100, 10, 1e-3, and 1e-2, respectively. The Dice coefficient was used for evaluating the segmentations.

\begin{table}[!t]
  \caption{Lesion segmentation Dice on lower torso}
  \centering
  \resizebox{\columnwidth}{!}{
  \begin{tabular}{@{}l|ccc@{}}
    \midrule
    Method & Healthy (71) & Disease (31) & Overall (102)\\
    \midrule
    SegOnly        & 0.0007 $\pm$ 0.0026         & 0.1864 $\pm$ 0.2474                      & 0.0572 $\pm$ 0.1598  \\
    SegRecon        & 0.0013 $\pm$ 0.0048         & 0.1847 $\pm$ 0.2474                      & 0.0570 $\pm$ 0.1593 \\
    SegReconHealthy & 0.0008 $\pm$ 0.0012         & 0.1791 $\pm$ 0.2403                      & 0.0550 $\pm$ 0.1547 \\
    \methodname     & \textbf{0.7174 $\pm$ 0.4200} & \textbf{0.5153 $\pm$ 0.2843} &  \textbf{0.6560 $\pm$ 0.3937}\\
    \bottomrule
  \end{tabular}}
  \label{tab:lowerdice}
\end{table}

\noindent \textbf{Baseline and Ablation Experiments:} To investigate whether disentanglement of healthy and disease features can improve lesion segmentation in PET images, \methodname is compared to baseline non-disentanglement variations that use 3D UNet: (i) SegOnly performs segmentation only, learning only disease features, (ii) SegRecon performs simultaneous segmentation and reconstruction learning both disease and image features, and (iii) SegReconHealthy learns disease and healthy features by performing segmentation on all examples and reconstruction on only healthy examples, providing a simpler approach to learn disjoint healthy and disease features compared to disentanglement.

\section{Results}
\label{sec:results}
Table \ref{tab:lowerdice} presents the Dice coefficients (mean $\pm$ standard deviation) obtained on the test set for healthy, disease, and overall examples for each experiment. We observe that SegOnly, SegRecon, and SegReconHealthy have significantly lower values compared to \methodname in each set of examples.

\begin{figure}
    \centering
    \includegraphics[width=\columnwidth]{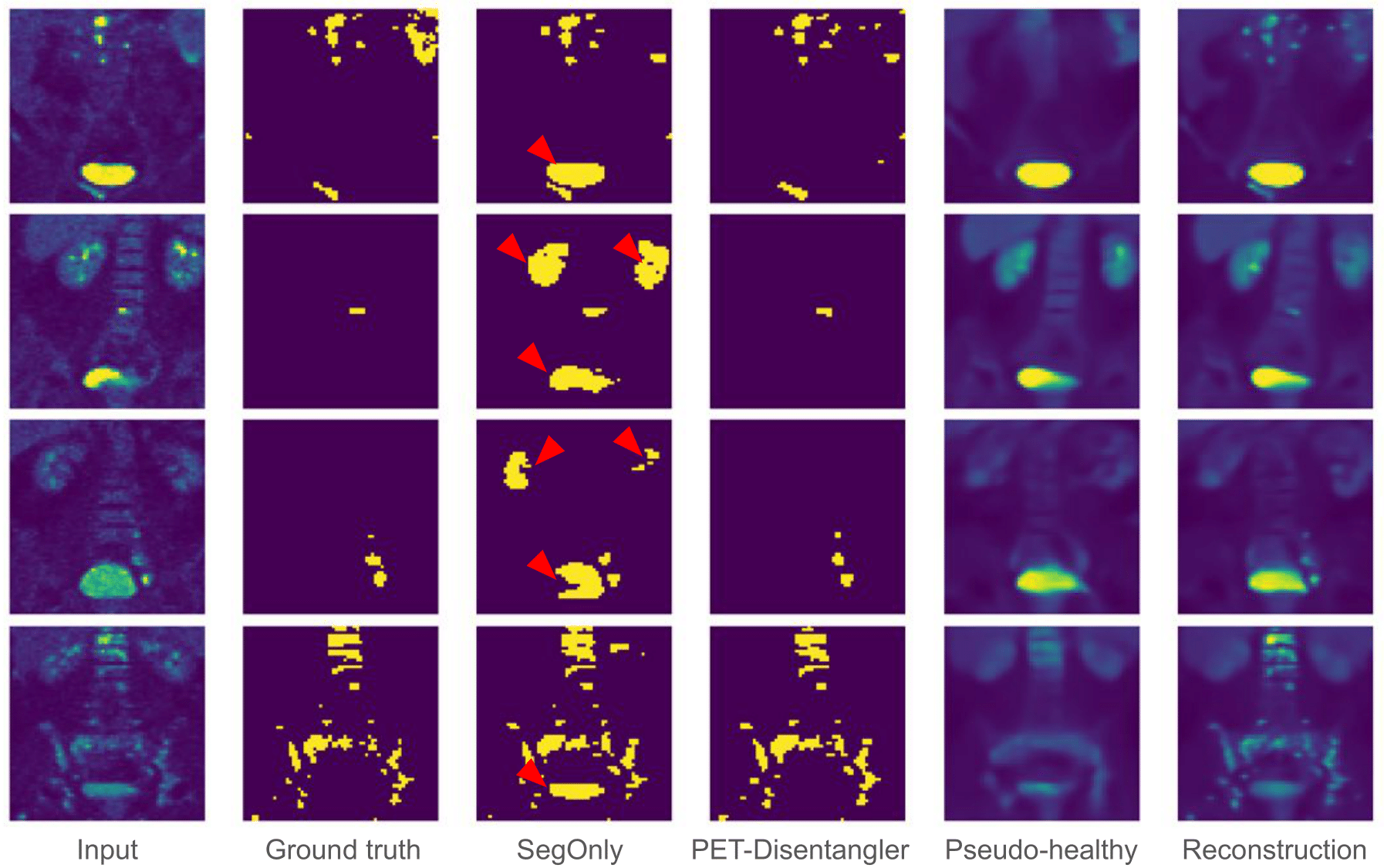}
    \caption{Comparison of \methodname to SegOnly, with red markers highlighting false positives for  healthy uptake patterns produced by SegOnly, which are not present in \methodname.
    }
    \label{fig:bladder_examples}
\end{figure}

Fig. \ref{fig:bladder_examples} presents segmentation prediction examples for SegOnly and \methodname, where we observe SegOnly consistently misclassifies the bladder as lesions given its high, yet normal, activity. Additionally, the second and third rows indicate that high intensity activity, that corresponds to healthy kidney uptake, is also incorrectly segmented as lesions. The red markers in Fig. \ref{fig:bladder_examples} highlight the incorrectly segmented normal activity by SegOnly.  In contrast, \methodname has learned the healthy anatomy component of the images and can identify these high uptake regions as healthy activity and focus on the remaining activity to accurately delineate the lesions.

Fig. \ref{fig:bladder_examples} visualizes \methodname-generated pseudo-healthy image and the corresponding image reconstruction. A comparison between these images shows an absence of lesion features, characterized by high intensity and abnormal uptake patterns, in the pseudo-healthy images. 
The pseudo-healthy images fill in these lesion regions with the expected ``healthy" appearance. These lesion features re-appear in the full reconstructions of the images, which are obtained by re-entangling the healthy and disease components via the SPADE blocks in the image decoder.

\section{Conclusions}
\label{sec:conclusions}
In this work, we proposed the use of image disentanglement for lesion segmentation by decomposing a PET image into healthy and disease features. Our method, \methodname, is a modified UNet architecture that learns to disentangle PET images into healthy and disease features in the latent space, producing segmentation predictions using the disease features, and re-entangles healthy and disease features to produce reconstructed images. A critic network is used to ensure the healthy features match the same distribution to prevent leakage of disease features. To the best of our knowledge, our work is the first to apply healthy and disease disentanglement to PET images and to extend this concept to 3D images.

Our evaluations on a whole-body FDG PET/CT dataset show \methodname greatly reduces the false positive segmentation of healthy uptake patterns, including but not limited to the bladder, compared to non-disentanglement methods by learning the healthy anatomy component. Future work can investigate the use of an additional modality (i.e., CT, MRI) in combination with PET. \\

\section{Compliance with Ethical Standards}
This research study was conducted retrospectively using human subject data made available in open access by TCIA \cite{gatidis2022whole}. Ethical approval was not required as confirmed by the license attached with the open access data.

\section{Acknowledgments}
\label{sec:acknowledgments}
This work was funded by the National Institutes of Health (NIH) / Canadian Institutes of Health Research (CIHR) Quantitative Imaging Network (QIN) (OQI-137993), and NSERC Discovery (RGPIN-06795). The authors are also grateful to the computational resources provided by NVIDIA Corporation, Digital Research Alliance of Canada, and SFU's Solar.

\bibliographystyle{IEEEbib}
\bibliography{refs_short}

\end{document}